\documentstyle[]{mn}
\newif\ifAMStwofonts

\def\simgt{\mathrel{\spose{\lower 3pt\hbox{$\sim$}}
        \raise 2.0pt\hbox{$>$}}}
\def\simlt{\mathrel{\spose{\lower 3pt\hbox{$\sim$}}\raise 2.0pt\hbox{$<$}}}

\ifoldfss
  \newcommand{\rmn}[1] {{\rm #1}}

  \ifCUPmtlplainloaded \else
    \NewTextAlphabet{textbfit} {cmbxti10} {}
    \NewTextAlphabet{textbfss} {cmssbx10} {}
    \NewMathAlphabet{mathbfit} {cmbxti10} {} 
    \NewMathAlphabet{mathbfss} {cmssbx10} {} 
  \fi
  \ifAMStwofonts
    \ifCUPmtlplainloaded \else
      \NewSymbolFont{upmath} {eurm10}
      \NewSymbolFont{AMSa} {msam10}
      \NewMathSymbol{\upi}     {0}{upmath}{19}
      \NewMathSymbol{\umu}     {0}{upmath}{16}
      \NewMathSymbol{\upartial}{0}{upmath}{40}
      \NewMathSymbol{\leqslant}{3}{AMSa}{36}
      \NewMathSymbol{\geqslant}{3}{AMSa}{3E}

    \fi
  \fi
\fi 

\ifnfssone
  \newmathalphabet{\mathit}
  \addtoversion{normal}{\mathit}{cmr}{m}{it}
  \addtoversion{bold}{\mathit}{cmr}{bx}{it}
  \newcommand{\rmn}[1] {\mathrm{#1}}

  \newmathalphabet{\mathbfit} 
  \addtoversion{normal}{\mathbfit}{cmr}{bx}{it}
  \addtoversion{bold}{\mathbfit}{cmr}{bx}{it}
  \newmathalphabet{\mathbfss} 
  \addtoversion{normal}{\mathbfss}{cmss}{bx}{n}
  \addtoversion{bold}{\mathbfss}{cmss}{bx}{n}
  \ifAMStwofonts
    \ifCUPmtlplainloaded \else
      \UseAMStwoboldmath
      \makeatletter
      \new@mathgroup\upmath@group
      \define@mathgroup\mv@normal\upmath@group{eur}{m}{n}
      \define@mathgroup\mv@bold\upmath@group{eur}{b}{n}
      \edef\UPM{\hexnumber\upmath@group}
      \new@mathgroup\amsa@group
      \define@mathgroup\mv@normal\amsa@group{msa}{m}{n}
      \define@mathgroup\mv@bold\amsa@group{msa}{m}{n}
      \edef\AMSa{\hexnumber\amsa@group}
      \makeatother
      \mathchardef\upi="0\UPM19
      \mathchardef\umu="0\UPM16
      \mathchardef\upartial="0\UPM40
      \mathchardef\leqslant="3\AMSa36
      \mathchardef\geqslant="3\AMSa3E
    \fi
  \fi
\fi 

\ifnfsstwo
  \newcommand{\rmn}[1] {\mathrm{#1}}

  \DeclareMathAlphabet{\mathbfit}{OT1}{cmr}{bx}{it}
  \SetMathAlphabet\mathbfit{bold}{OT1}{cmr}{bx}{it}
  \DeclareMathAlphabet{\mathbfss}{OT1}{cmss}{bx}{n}
  \SetMathAlphabet\mathbfss{bold}{OT1}{cmss}{bx}{n}
  \ifAMStwofonts
    \ifCUPmtlplainloaded \else
      \DeclareSymbolFont{UPM}{U}{eur}{m}{n}
      \SetSymbolFont{UPM}{bold}{U}{eur}{b}{n}
      \DeclareSymbolFont{AMSa}{U}{msa}{m}{n}
      \DeclareMathSymbol{\upi}{0}{UPM}{"19}
      \DeclareMathSymbol{\umu}{0}{UPM}{"16}
      \DeclareMathSymbol{\upartial}{0}{UPM}{"40}
      \DeclareMathSymbol{\leqslant}{3}{AMSa}{"36}
      \DeclareMathSymbol{\geqslant}{3}{AMSa}{"3E}
    \fi
  \fi
\fi 

\ifCUPmtlplainloaded \else
  \ifAMStwofonts \else 
    \def\upi{\pi}
    \def\umu{\mu}
    \def\upartial{\partial}
  \fi
\fi

\title[Gravitational Microlensing of Gamma Ray Bursts]
  {Gravitational Microlensing of Gamma Ray Bursts at Medium Optical Depth}
\author[J. S. B. Wyithe \& E. L. Turner]
  {J.~S.~B.~Wyithe$^{1,2}$, 
  E. L.~Turner$^2$\\
  $^1$ School of Physics, University of Melbourne, Parkville, Vic, 3052, Australia\\
  $^2$ Princeton University Observatory, Peyton Hall, Princeton, NJ 08544, USA\\ 
 Email: swyithe@astro.Princeton.edu, elt@astro.Princeton.edu }
\date{Accepted Received}
\pagerange{\pageref{firstpage}--\pageref{lastpage}}
\pubyear{1999}

\def\LaTeX{L\kern-.36em\raise.3ex\hbox{a}\kern-.15em
    T\kern-.1667em\lower.7ex\hbox{E}\kern-.125emX}

\begin{document}

\label{firstpage}

\maketitle

\begin{abstract}

Gravitational lensing of a gamma ray burst (GRB) by a single point mass will produce a second, delayed signal. Several authors have discussed using microlensed GRBs to probe a possible cosmological population of compact objects. We analyse a closely related phenomenon; the effect of microlensing by low to medium optical depth in compact objects on the averaged observed light-curve of a sample of GRBs. We discuss the cumulative measured flux as a function of time resulting from delays due to microlensing by cosmological compact objects. The time-scale and curvature of this function describe unique values for the compact object mass and optical depth. For GRBs with durations larger than the detector resolution, limits could be placed on the mass and optical depth of cosmological compact objects. The method does not rely on the separation of lensed bursts from those which are spatially coincident.  

\end{abstract}


\section{Introduction}

Press \& Gunn (1973) were the first to propose that a cosmological abundance of dark compact objects could be detected by gravitational lensing of more distant sources. If a point mass lies along the observer-source line of sight, the relative motion between the lens, source and observer produces a change in the magnifications of two lensed images. The presence of foreground compact objects is therefore detected through a change in the observed flux of the background source (termed microlensing).
 This effect has been used successfully in the search for compact objects in the halo of the Milky Way galaxy (e.g. Alcock et al. 2000). Also, microlensing due to stars in a galaxy at moderate redshift has been observed in the gravitationally lensed quasar Q2237+0305 (Irwin et al. 1989; Corrigan et al. 1991). The short durations of gamma ray bursts (GRBs) offer an alternative way to study microlensing and hence search for a cosmological population of compact objects, through observation of repeating bursts.

In this paper we assume GRBs to be at cosmological distances (Paczynski 1995), and therefore probable sources for gravitational lensing (Turner, Ostriker \& Gott 1984). Paczynski (1986, 1987) noted that while multiple images of a single GRB cannot be angularly resolved by present day detectors, their relative delay may be longer than the burst duration so that the lensed images could be resolved temporally (a pair of lensed GRB images that are produced by a single mass $(M_{CO})$ have a relative time delay that is $\Delta t\sim 50 sec\times(M_{CO}/10^{6}M_{\odot})$ (Mao 1992)). Microlensing of a GRB by a compact object is therefore observed as a GRB that repeats. The utility of GRBs to explore cosmological dark matter, as well as the possible inference of properties of the GRB population itself was discussed in detail by Blaes \& Webster (1992). The short event duration, as well as the transparency of the universe to gamma rays make GRBs ideal probes of dark matter in the form of compact objects over a wide range of masses.

Microlensing of existing and potential catalogues of GRBs have been used to discuss the cosmological abundance of compact objects.
Marani et al. (1999) use non-detections of lensed images from the BATSE and Ulysses catalogues to set conservative limits on dark compact objects with masses between $10^{-16}$ and $10^{-7}M_{\odot}$. Also a universe proposed by Gnedin \& Ostriker (1992) with $M_{CO}\sim10^{6.5}M_{\odot}$ and $\Omega_{M}=\Omega_{CO}=0.15$ was ruled out at a confidence level of 90\%. This scenario had been previously investigated in detail, using a 3-D lensing code by Mao (1993). He found results that did not depart significantly from those obtained by a single-screen approximation. Mao (1992) and Grossman \& Nowak (1994) estimate that the waiting time for one lensed pair due to an intervening galaxy to be observed in the BATSE catalogue is between one and 10 years, and conclude that it is not certain that such a lensed pair will be found by BATSE. 

As in quasar lensing, the detection of a lensed pair is separated from spatially coincident events by the comparison of light-curves and spectra. However the presence of noise and the faintness of images will alter the light curves so that macro images appear dissimilar (Wambsganss 1993; Nowak \& Grossman 1994). In addition, if a foreground galaxy is responsible for the lensed pair then Williams \& Wijers (1997) find that microlensing by individual stars in that galaxy can smear out image light curves, further increasing the chance of their being mis-classified as two spatially coincident events. Light-curve similarity of a lensed pair requires the source to be much smaller than the Einstein ring radius of the lens (projected into the source plane), a condition likely to be filled in the case of GRBs due to the short event duration for all but the smallest lenses. However an anisotropic source (eg. a beamed source) is effectively viewed at two different angles in the two images. This raises the possibility that the time variation of the two images may be different (eg. Babul, Paczynski \& Spergel 1987), although Blaes \& Webster (1992) find that a source should be isotropic to lensing provided that the beaming is not too strong. Paczynski (1987) noted that the potential difficulties in identifying lensed pairs will make confirmation of the cosmological origin of the GRBs difficult through a lensing argument. On the other hand, if we assume that GRBs have a cosmological origin, the mis-classification of events will lead to an underestimate of a compact object population.

The microlensing effect of a cluster of objects on a GRB was first discussed by Paczynski (1987). He found that a single instantaneous burst lensed by a screen of objects such as that produced by a galactic halo or a cluster of galaxies results in the observation of many repeats of the original burst separated by different delays. Examples of the light-curves produced can be found in Paczynski (1987) and Williams \& Wijers (1997). At moderate optical depths the first images have magnifications that are not correlated with their delay, however the faint images arriving later have observed fluxes that decrease monotonically with time. Williams \& Wijers (1997) investigated the flux weighted $rms$ of the delay of clusters of microimages and found this to be a sensitive function of optical depth and shear. The temporal spread of the cluster of microimages was shown to be related to the longest side of the area of the image plane containing microimages that account for most of the macroimage flux. This scale length is considerably larger than that of the separation of a lensed pair produced by a single microlens, and so the spread in the flux weighted arrival time is also considerably larger than the relative delay between images produced by a single mass. This phenomenon allows smaller masses to be probed modulo a detector resolution and intrinsic GRB duration.

 A non-negligible fraction of the baryon content of the universe may exist in stellar remnant form (eg. Kerins \& Carr 1994). In addition, Galactic microlensing searches (e.g. Alcock et al. 2000) provide evidence for a significant fraction of the Galactic halo mass being in the form of stellar mass compact objects, an interpretation which is supported by observations of high proper motion, cold white dwarfs (Ibata et al. 2000; Hodgkin et al. 2000). While intervening galaxies produce relatively large delays, GRBs with micro-second durations or variability would resolve microlensing by cosmological stellar mass objects (Nemiroff, Norris, Bonnell \& Marani 1998). Nemiroff et al. (1998) note that there is no known fundamental reason for the non-existence of micro-second GRBs, and find that a significant rate of these might exist and go undetected by current telescopes. They find that at a given flux level there may be more GRBs with durations between 1 and 2 milliseconds than between 8.192 and 16.384 seconds (BATSE duration bin), and an abundance of microsecond spikes that is an order of magnitude greater. However, Nemiroff et al. (1998) also note that the converse possibility of short duration spikes not existing at all is equally consistent with the analysis of current data.

We propose that if short duration bursts were to exist then they could be used to probe the cosmological stellar mass compact object population without resolving and identifying individual micro-images. Also, in the absence of short spikes, the light-curves of GRBs can place upper limits on typical compact object mass and optical depth. Rather than considering the probability of lensing by a single mass on individual GRBs, we consider the microlensing effect on an ensemble of GRB light curves of a collection of point masses at low to moderate optical depth. For a sample of microlensed spike GRBs we look at the average value of the cumulative flux as a function of time. 

This paper presents calculations corresponding to the microlensing effect of populations of stellar mass objects on the observed light-curves of hypothetical, very-short duration GRBs. However, the results are also applicable to the presently known population of GRBs, with durations of 10s of seconds as probes of compact objects having masses of $\sim10^{6}M_{\odot}$. The relevant scaling is pointed out where appropriate.  

The remainder of the paper is presented in 3 parts. Section \ref{models} describes the microlensing models and section \ref{method} proposes a new method for analysing microlensing in GRBs. A brief summary is presented in section \ref{disc}.

\section{Microlensing models}
\label{models}

\begin{figure*}
\vspace*{110mm}
\includegraphics{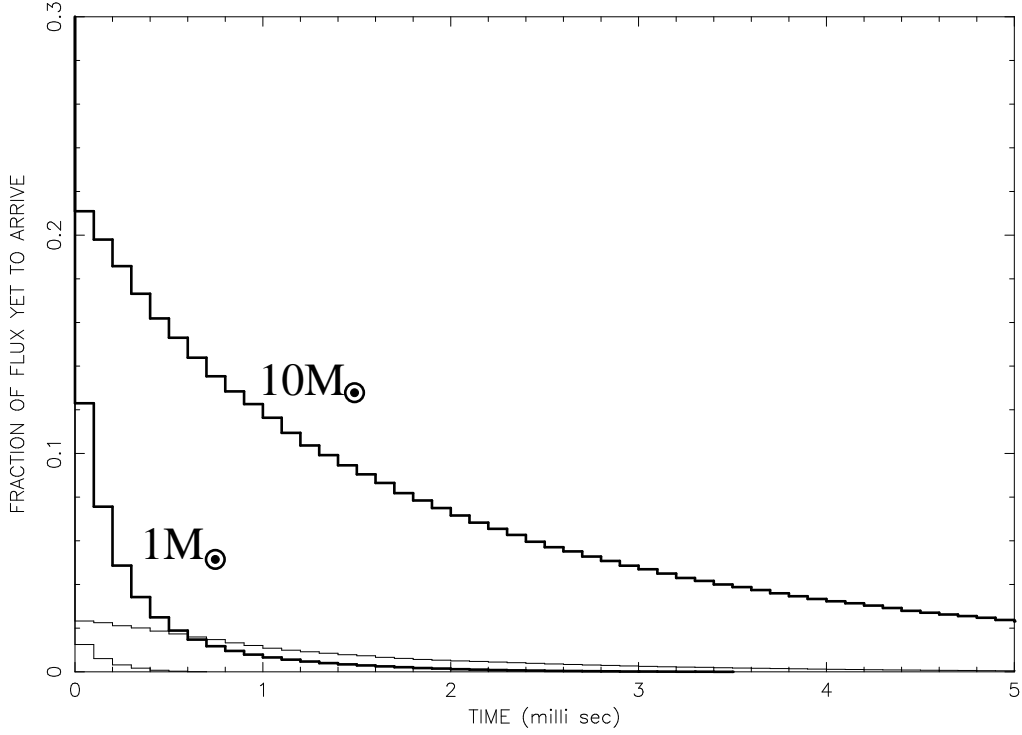}
\caption{\label{mass_dependence_2}The fraction of flux that is yet to arrive as a function of time. The cases shown are for mean masses of 1$M_{\odot}$ and 10$M_{\odot}$ for optical depths of $\kappa=$0.025 (thin lines) and $\kappa=$0.250 (thick lines).}
\end{figure*}

\begin{figure*}
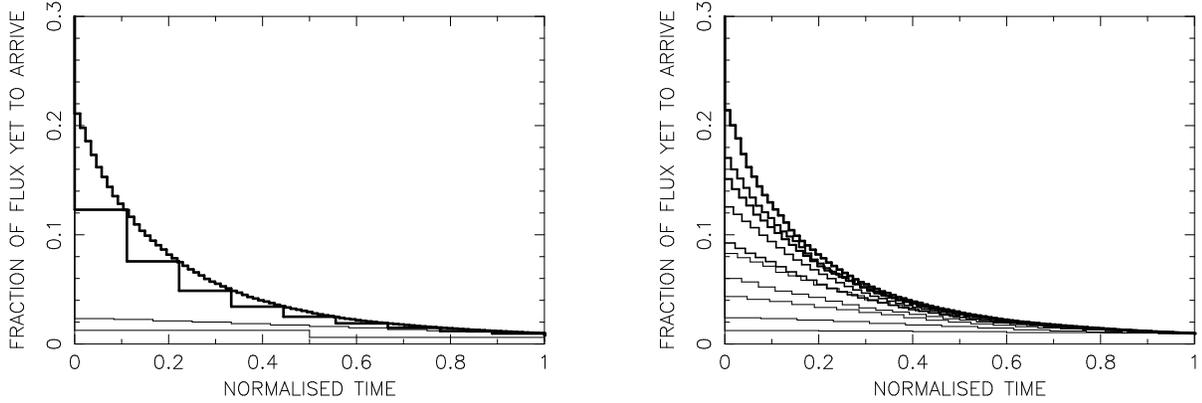

\vspace*{60mm}
\includegraphics{fig2a.ps}
\includegraphics{fig2b.ps}
\caption{\label{opt_dependence}The fraction of flux that is yet to arrive as a function of time normalised to $T_{99}$. Left: The curves correspond to the cases shown in Fig. \ref{mass_dependence_2}. Right: The curves correspond to optical depths $\kappa=$0.025, 0.050, 0.075, 0.100, 0.125, 0.150, 0.175, 0.200, 0.225, 0.250. (thicker lines denote larger $\kappa$) at a mean mass of 10$M_{\odot}$.}
\end{figure*}

As a crude approximation to the lensing effect of compact objects distributed along the line of sight, we assume a screen of point masses distributed randomly in a disc. The model does not include a continuous matter distribution.
 We use standard notation for gravitational lensing. The Einstein radius of a microlens in the image plane is denoted by $\xi_{o}$ and when projected into the source plane by $\eta_{0} $. The normalised convergence or optical depth is denoted by $\kappa$.
 The normalised lens equation for a field of point masses is
\begin{equation}
\vec{y}= \left( \begin{array}{cc}
         1-\gamma & 0 \\
        0 & 1+\gamma 
            \end{array} \right)\vec{x} + \sum_{j=0}^{N_{*}}m^{j}\frac{(\vec{x}^{j}-\vec{x})}{|\vec{x}^{j}-\vec{x}|^{2}}
\label{lens_map} 
\end{equation}
Here $\vec{x}$ and $\vec{y}$ are the normalised image and source positions respectively, $\gamma$ is the applied shear, and the $\vec{x_{i}}^{j}$ and $m^{j}$ are the normalised positions and masses of the individual microlenses. To construct a microlensed light-curve for an instantaneous microlensed GRB, Eqn. \ref{lens_map} is solved for the micro-image positions at many points along a predefined source line through the inversion technique of Lewis et al. (1993) and Witt (1993). The time delay and magnification are then determined for each micro-image $i$, given by
\begin{eqnarray}
\nonumber
\Delta T_{i}&=&\\
&&\hspace{-15mm}\frac{\xi_{o}^{2}}{c} \frac{D_{s}}{D_{d}D_{ds}}(1+z_{d})\left(\frac{(\vec{x}_{i}-\vec{y})^{2}}{2}-\sum_{j=0}^{N_{*}}m^{j}ln(|\vec{x}_{i}-\vec{x}^{j}|)\right)
\end{eqnarray}
and
\begin{eqnarray}
\mu_i&=&\frac{1}{|det\,A(\vec{x}_{i})|}\\\nonumber
 &&\hspace{-12mm}  {\rmn where} \hspace{5mm}  det\,A(\vec{x}_{i}) = \frac{\partial y_1}{\partial x_1}|_{\vec{x}_{i}}\frac{\partial y_2}{\partial x_2}|_{\vec{x}_{i}}-\left(\frac{\partial y_2}{\partial x_1}|_{\vec{x}_{i}}\right)^2.
\end{eqnarray}
 Here $D_{s},D_{d}$ and $D_{ds}$ are the angular diameter distance between the observer and the source, the observer and the lens, and the lens and source. $c$ is the speed of light and $z_{d}$ the lens redshift. We note that the delay is proportional to ($1+z_{d}$), but is not explicitly dependent on the source redshift. Also, due to the dependence on $\xi_{o}^{2}$, $\Delta T$ is proportional to the mean mass.

 At a sufficiently large angle from the point source, there is a low magnification image located very close to each point mass. All solutions of the lens equation must therefore be found in a region that contains a sufficient percentage of the total macroimage flux. The region of the lens plane in which image solutions need to be found to ensure that $99.9\%$ of the total macro-image flux is recovered from all points on the source line is known as the shooting region. The number of stars in the region about any point which collects 99.9\% of the macro-image flux was calculated by  Katz, Balbus \& Paczynski (1986) and is given by:
\begin{equation}
N_{*}=300\frac{\langle m^{2}\rangle}{\langle m \rangle ^{2}}\frac{\kappa^{2}}{|(1-\kappa)^{2}-\gamma^{2}|}.
\end{equation}
In the absence of shear ($\gamma=0)$ these stars are distributed in a disc with a radius $R_{sd}=\sqrt{\kappa/(N_{*}\times \langle m \rangle)}$.
The shooting region is defined by the union of these discs centred on the points $(x_{1}=y_{1}/(1-\kappa),x_{2}=y_{2}/(1-\kappa))$ corresponding to all parts of the source line. The minimum number of stars ($N_{min}$) required for the model are contained in a disc having a radius $R_{*}$ which covers the shooting region. We have used 500 stars in each of our models which is larger than $N_{min}$ in all cases. We assume that the source is stationary with respect to the microlenses for the duration of the microlensed light-curve. We find the light curves for 10 bursts per microlens field distributed along a source track of length 10$\eta_{o}$.
At optical depths between $\kappa=0.025$ and 0.25 (at intervals of 0.025) we compute the light curves for $10^4$ bursts.

The most probable scenario for lensing has the lens lying at a distance that is a reasonable fraction of that of the source (Turner, Ostriker \& Gott 1984). In addition, the distribution of optical depth with redshift is reasonably sharply peaked. With this in mind we stress that our calculation uses a single screen to approximate the lensing effect of all masses along the line of sight. As an example we place the population of model GRBs at a redshift of $z_{s}=1$ and a screen of compact objects at a redshift of $z_{d}=\frac{2}{3}$. For simplicity we assume that all compact objects have a unique mass $m$.

\section{Method of analysis}
\label{method}

\begin{figure}
\vspace*{90mm}
\includegraphics{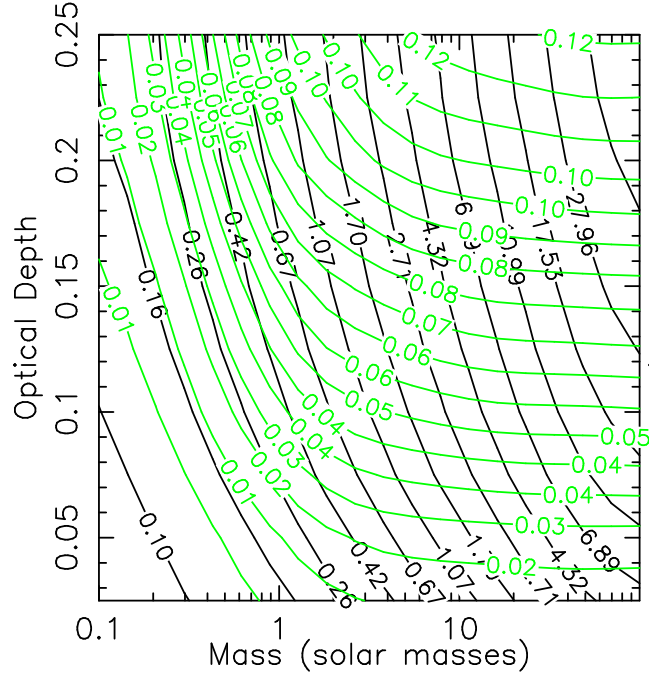}
\caption{\label{contour} Contour map showing $T_{0.99}$ in milliseconds (dark lines) and $F_{\Delta T_{norm}=0.1}$ (light lines) as a function of optical depth and mass.}
\end{figure}

In this section we consider the analysis of a hypothetical sample of spike GRBs that have an intrinsic duration smaller than 0.1 milliseconds which we take as the detector binsize. From our sample of model microlensed light curves we find the cumulative flux (summed over all located microimages) that has arrived by the end of each bin. One minus the cumulative flux divided by the total flux gives the fraction of total flux that is yet too arrive $(F_{i}(\Delta T))$ for each burst $i$ as a function of time. These curves are averaged over many bursts:
\begin{equation}
\label{F}
F(\Delta T)\equiv\sum_{i=0}^{N_{bursts}}F_{i}(\Delta T)/N_{bursts}
\end{equation}
Eqn. \ref{F} describes the average behaviour of the arrival time of lensed flux for a set of microlensing parameters $\kappa$ and $m$. Note the assumption that all lines-of-sight have the same average optical depth $\kappa$. From the work on clusters of GRB microimages by Paczynski (1987) and Williams \& Wijers (1997) we expect $F(\Delta T)$ to have the following characteristics. Since the microlensing spread of the GRB is related to the size of the shooting region, which is proportional to the Einstein radius of the microlenses and therefore to $\sqrt{m}$, $\Delta T(m)$ such that $F(\Delta T)=const$ is proportional to $m$. Secondly, at larger optical depths there are more microimages with large magnification. A given fraction of microlensed flux therefore arrives later at higher optical depths and so $F(\Delta T)$ is larger for increased $\kappa$. Note that when in a region where $(1-\kappa)^2-\gamma^2\sim0$, the magnification becomes very large, and flux from the burst will be observed at a comparable level for a time similar to the geometric delay for a trajectory at the edge of the region.

Fig. \ref{mass_dependence_2} shows the fraction of flux that is yet to arrive as a function of time. Cases are shown for mean masses of 1$M_{\odot}$ and 10$M_{\odot}$, at optical depths of $\kappa=$0.025 (thin lines) and $\kappa=$0.250 (thick lines). At each optical depth, the larger mass produces longer microimage delays, causing a given fraction of flux to arrive later on average. In addition, at a fixed mass a given fraction of flux arrives later at larger optical depths producing a curve that has a smaller initial drop in the first bin followed by a decline that is more rapid than that at smaller optical depth. However at higher $\kappa$, $F(\Delta T)$ is larger for all $\Delta T$.

We define $T_{0.99}$ as the time $\Delta T$ at which 99\% of the total microlensed flux from the GRB has arrived. For binsizes that are small with respect to the microlensed spread of the event, $T_{0.99}$ scales linearly with microlens mass. $T_{0.99}$ is also a function of $\kappa$ and provides a natural scaling unit for time. The left hand panel in Fig. \ref{opt_dependence} re-displays the curves from Fig. \ref{mass_dependence_2} with time normalised by $T_{0.99}$ in each case ($F(\Delta T_{norm})$ where $\Delta T_{norm} = \Delta T/T_{0.99}$). At the end of each bin the normalised curves are independent of mass. The curvature therefore provides a probe of the optical depth. The right hand panel of Fig. \ref{opt_dependence} displays $F(\Delta T_{norm})$ for optical depths between $\kappa=$0.025 and 0.250 at a microlens mass of $10M_{\odot}$. We suggest the value of $F(\Delta T_{norm}=0.1)$ as an indicator of optical depth. 

For values of $\kappa$ and $m$ in the ranges $0.025\la \kappa\la0.25$ and $0.1\la m\la100$, we made 1000 simulations of $F(\Delta T)$ (each $F(\Delta T)$ was calculated using 100 model microlensed GRB light-curves). Fig. \ref{contour} shows average values of the quantities $T_{0.99}$ (dark contours) and $F(\Delta T_{norm}=0.1)$ (light contours) over a range of optical depths and microlens masses. In regions combining small values of optical depth with microlens masses $m<1M_{\odot}$, the contours are nearly parallel, however for most of the parameter space, values of $T_{0.99}$ and $F(\Delta T_{norm}=0.1)$ describe a unique set of $\kappa$ and $m$. For samples containing 100 GRBs the variance in values of $T_{0.99}$ and $F(\Delta T_{norm}=0.1)$ are an order of magnitude lower than the corresponding mean. 

We have demonstrated that a sample of GRBs could be used to measure the quantities $T_{0.99}$ and $F(\Delta T_{norm}=0.1)$, and that this combination corresponds to measurements of $\kappa$ and $m$. However it will not be known whether the observed spread in the arrival time of GRB flux is due solely to microlensing effects or whether it is intrinsic to the source (unless the individual micro-images are resolved). $T_{0.99}$ and $F(\Delta T_{norm}=0.1)$ will therefore be upper bounds on the values for instantaneous bursts, and so will exclude a region of mass - optical depth parameter space rather than measure probable values. In addition, we note that Fig. \ref{contour} presents results covering only 3 orders of magnitude in mass. However the binsize, mass and $T_{0.99}$ are all linearly related, allowing scaling of the mass and $T_{0.99}$ by the binsize divided by 0.1 milliseconds. Estimates of upper limits for $T_{0.99}$ and $F(\Delta T_{norm}=0.1)$ which will probe $\sim10^6M_{\odot}$ compact objects are therefore possible using current data with GRB durations of $\sim 10$ seconds.

\begin{figure*}
\vspace*{95mm}
\includegraphics{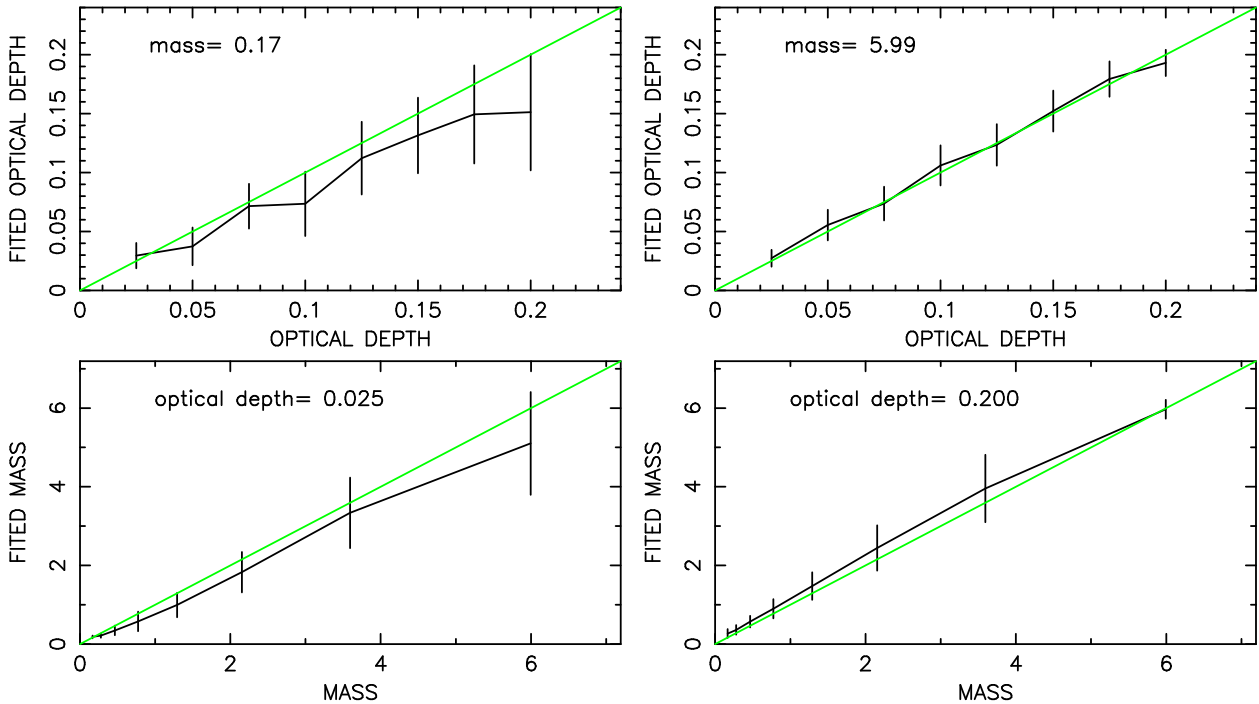}
\caption{\label{fitted_vals} Plots of known $\kappa$ vs. retrieved $\kappa$ at two masses (top), and known $m$ vs retrieved $m$ at two values for $\kappa$ (bottom). The error bars shown are the variance in the retrieved values. The sample contained 1000 simulations containing 100 bursts each.}
\end{figure*}

The approach described requires that the source be integrated for long enough to be sure that the entire event flux (or some appropriately large fraction) has been accumulated. We now describe an alternative approach that does not require measurement of the total flux. Fig. \ref{opt_dependence} demonstrates that the shape of $F(\Delta T)$ is unique for each combination of $m$ and $\kappa$. For example, a curve produced by a sample of GRBs lensed by a low optical depth of high masses may have the same value $F(\Delta T=T_{0.99})$ as a sample lensed by smaller masses having a higher optical depth. However at all $\Delta T<T_{0.99}$ the former will have a smaller value of $F(\Delta T)$. We have produced the average function $F_{av}(\Delta T)$ for masses and optical depths covering the parameter space shown in Fig \ref{contour}. At each combination of mass and optical depth ($m_{true}$ and $\kappa_{true}$), we make 1000 mock observations of $F_i(\Delta T)$ each calculated from 100 GRBs. We look for the values of $m$ and $\kappa$ such that $F_{av}(\Delta T)$ best fits each mock observation $F_i(\Delta T)$. Our criteria for best fit is to minimise the value
\begin{equation}
D=max(|F_i-F_{av}|).
\end{equation}
This procedure provides a likelihood for measuring $m$ and $\kappa$ given the true values $m_{true}$ and $\kappa_{true}$: 
\begin{equation}
p_{lh}(m,\kappa|m_{true},\kappa_{true})
\end{equation}
The construction of $F_{obs}(\Delta T)$ for an observed set of GRBs then provides estimates of $m$ and $\kappa$ through application of Bayes' theorem: 
\begin{equation}
p(m,\kappa) = N\int p_{lh}(m,\kappa|m',\kappa')\,p_{prior}(m',\kappa')dm'd\kappa',
\end{equation}
where $p_{prior}(m,\kappa)$ is the assumed prior probability for $m$ and $\kappa$, and $N$ is a normalising constant.

To demonstrate the statistical uncertainty we have plotted in Fig. \ref{fitted_vals} the mean and variance of retrieved values of $m$ and $\kappa$ verses their assumed values ($\kappa_{true},m_{true}$) at fixed mass and optical depth respectively (ie. sections through $p_{lh}$). As in the previous calculation, the figure demonstrates that a sample of 100 GRBs is sufficient to obtain a consistent result. The systematic bias (which is smaller than the statistical uncertainty) in the recovered values is due to the finite grid of $m$ and $\kappa$ over which $F_{av}$ and $p_{lh}$ are computed. This calculation would obtain upper limits on the true values of $m$ and $\kappa$ due to possible intrinsic spread of the GRB. Also, the binsize and mass are linearly related allowing scaling of the case presented here.

\section{Discussion}
\label{disc}

If gamma ray bursts have a cosmological origin then they provide a unique opportunity to study the cosmological abundance of compact objects through the identification of lensed pairs. However there are several mechanisms that may lead to the mis-classification of lensed pairs as spatially coincident but independent bursts. In addition, the identification of a lensed pair may be ambiguous if the relative delay is smaller than the event duration. Upper limits on the contribution to the mass density of compact objects determined from a lack of lensed pairs may therefore be underestimated.

We have demonstrated that simultaneous upper limits on the average optical depth and mass of compact objects can be estimated by measuring the average cumulative flux as a function of time for a collection of bursts and without resolving the individual lensed images. By selecting sub-samples of bursts, specific populations of compact objects could be examined. For example, a collection of bursts that are spatially coincident with fore ground galaxies (like GRB 971214 (Kulkarni et al. 1998; Diercks et al. 1999)) will probe the compact object population in galactic halos. In particular, if a population of short duration (micro-second) cosmological gamma ray bursts are identified, limits will be placed on the abundance of stellar mass compact objects.

\section*{Acknowledgements}

This work was supported by NSF grant AST98-02802. JSBW acknowledges the support of an Australian Postgraduate award and a Melbourne University Overseas Research Experience Award. We would like to thank Rachel Webster for numerous helpful discussions.

\label{lastpage}

\end{document}